\documentclass[preprint2]{aastex}

\slugcomment{}

\shorttitle{Coupling between radial and torsional oscillations}
\shortauthors{Zaqarashvili and Belvedere}

\begin{document}


\title{Coupling between radial and torsional oscillations\\ in a
magnetized plasma and possible stellar applications}


\author{T.V. Zaqarashvili}
\affil{Georgian National Astrophysical Observatory (Abastumani
Astrophysical Observatory), Al. Kazbegi ave. 2a, 0160 Tbilisi,
Georgia} \email{temury@genao.org}

\and

\author{G. Belvedere}
\affil{Dipartimento di Fisica e Astronomia, Sezione Astrofisica,
Universit{\'a} di Catania, Via S.Sofia 78, I-95123 Catania, Italy}
\email{gbelvedere@ct.astro.it}



\begin{abstract}
The weakly non-linear interaction of slow magnetoacoustic and
torsional Alfv{\'e}n waves propagating along an unperturbed poloidal
magnetic field in a stellar interior is studied in the high plasma
$\beta$ limit. It is shown that slow magnetoacoustic waves
parametrically drive torsional modes with the half frequency and
wave number. Therefore global slow magnetoacoustic waves which have
a radial velocity polarization, may amplify the torsional
oscillations. Possible applications of this mechanism to the Sun and
binary stars are briefly discussed.
\end{abstract}



\keywords{Stars:magnetic fields -- Stars:oscillations --
Magnetohydrodynamics (MHD) -- Waves }


\section{Introduction}

Although Wal{\'e}n's original suggestion \citep{wal} that torsional
azimuthal Alfv{\'e}n oscillations may exist in stellar interiors,
the energy source sustaining these oscillations is not yet known
\citep{ros}. Without an energy source, the oscillations are believed
to be damped by phase mixing \citep{cha} or transport of magnetic
flux towards the surface. Several different mechanisms have been
suggested from time to time to support torsional oscillations
\citep{la,ma,lan,ga}, however none of them may account for the
energy source. The main problem is that torsional Alfv{\'e}n waves
are pure electromagnetic oscillations in magnetohydrodynamics and
one needs to have an azimuthal force which may support them against
damping. The absence of such a force suppressed the interest in
these oscillations as a possible mechanism for stellar activity
cycles. However no one paid attention to the fact that it is not
necessary to have a direct driving force for the amplification of
the oscillations. The oscillation can also be amplified through the
periodical variation of the system parameters and the process is
called parametric resonance. The well-known example of such system
is the mathematical pendulum with periodically varying length. When
the frequency of the parameter variation is twice the frequency of
the system oscillation, then the oscillations will exponentially
grow in time. The phase speed of torsional Alfv{\'e}n waves depends
on the magnetic field and medium density. So if some external force
(or oscillation) causes a periodical variation of either quantities
(or both together) then the Alfv{\'e}n waves will be amplified. The
pioneering work of \citet{zaq} suggested that the periodical density
variation caused by sound waves in the high plasma $\beta$ limit may
amplify the shear Alfv{\'e}n waves. Then, \citet{zaq1} showed the
resonant energy transfer from fast magneto-acoustic waves
propagating across the unperturbed magnetic field to Alfv{\'e}n
waves propagating along the field. The energy transfer is relative
to the velocity component of resonant Alfv{\'e}n waves normal to
both magnetic field and direction of propagation of the
magnetoacoustic waves. In both papers the temporal behaviour of
Alfv{\'e}n waves was found to be governed by Mathieu equation. Thus
the frequency of resonant Alfv{\'e}n waves was the half frequency of
the driving compressible oscillations. The natural astrophysical
suggestion was that stellar pulsation in the fundamental mode may
amplify torsional Alfv{\'e}n waves in a seed magnetic field
\citep{zaq3}. The process can be of importance in stellar
chromospheres and coronas as the Alfv{\'e}n waves propagate upwards
carrying energy and momentum and so may contribute to chromospheric
and coronal heating. Also the waves may propagate up to the stellar
wind and recently \citet{zaq5} suggested them as a source for long
period Alfv{\'e}n oscillations (with periods of few hours) observed
by ULYSSES in the solar wind. On another hand, the fundamental
frequency of stellar pulsation is too high for resonant torsional
Alfv{\'e}n waves to give rise to large-scale torsional oscillations
in a stellar interior. Therefore one needs a longer period pulsation
in order to suggest them as an energy source for torsional
oscillations.

Here we suggest that the slow magnetoacoustic waves propagating
along the magnetic field in the high plasma $\beta$ limit are
nonlinearly coupled to the torsional Alfv{\'e}n waves propagating in
the same direction. The velocity polarization of slow
magnetoacoustic waves propagating along a straight uniform magnetic
field is along the radial direction in the case of a cylindrical
coordinate system (the waves have a velocity component along the
magnetic field as well). Therefore fixing the boundaries along the
axis of the cylindrical system originates a standing pattern of
waves leading to the pulsation of the magnetic cylinder with a
spatial scale which corresponds to the first or second eigenmode. In
the case of stars with a poloidal magnetic field in the interior,
the fundamental mode of slow magnetoacoustic waves has a very low
frequency due to the small phase speed (note, that the slow
magnetoacoustic waves propagate with the Alfv{\'e}n speed in high
$\beta$ limit). So it will lead to long period stellar pulsation.
Since the velocity of the slow magnetoacoustic waves has the radial
direction, any force (gravity, radiation pressure etc) may support
this oscillation. If the oscillation transfers energy to torsional
Alfv{\'e}n waves, then they will lead to the set up of torsional
oscillations, as their wavelength will comparable to the stellar
radius. So the torsional oscillations will gain energy from the
radial pulsation and thus can be sustained against damping. Here we
consider a simple case of wave coupling in an uniform magnetic
cylinder, which can be accounted as the simplest model of a
non-rotating star with a poloidal magnetic field.





\section{Equations and mathematical developments}

We consider a non-turbulent medium with zero viscosity, infinite
conductivity and negligible displacement current, which can be
described by the magnetohydrodynamic (MHD) equations:
\begin{equation}
{{{\partial \bf B}}\over {\partial t}}={\nabla}{\times}({\bf
V}{\times}{B}),
\end{equation}
\begin{equation}
{\rho}{{{\partial \bf V}}\over {\partial t}} + {\rho}({\bf
V}{\cdot}{\nabla}) {\bf V} = - {\nabla}P + {1\over
{\mu}}{({\nabla}{\times}{\bf B}){\times}{\bf B}},
\end{equation}
\begin{equation}
{{{\partial {\rho}}}\over {\partial t}} + {\nabla}({\rho}{\bf V})=0,
\end{equation}
where $\rho$ is the medium density, $P$ is the pressure, ${\bf V}$
is the velocity, $\bf B$ is the magnetic field and $\mu$ is the
magnetic permeability. To this system we add the adiabatic equation
of state
\begin{equation}
P=p_0\left ( {{\rho}\over {\rho_0}}\right  )^{\gamma},
\end{equation}
where $p_0$ and $\rho_0$ are the unperturbed pressure and density,
while $\gamma$ is the ratio of specific heats. We adopt a
cylindrical coordinate system ($R$, $\phi$, $Z$) and for simplicity
consider only the axisymmetric problem, so everywhere ${{\partial
}/{\partial {\phi}}}=0$ is assumed. We suppose that an uniform
magnetic field is directed along the $Z$ axis so $B=(0,0,B_0)$.

The medium is supposed to be uniform, therefore the equilibrium is
satisfied automatically. It must be mentioned, however, that real
astrophysical situation is much more complicated. In the stellar
interior the gravity and the pressure gradient necessarily lead to
the non uniform spatial distribution of unperturbed physical
quantities (density, temperature and magnetic field). But in order
to understand the physics of the wave coupling we choose the
simplest uniform model (Fig.1), which considers an uniform magnetic
cylinder with an uniform density. Thus, gravity is neglected in the
momentum equation (note that almost all dynamo models also ignore
the gravity). Once the physical grounds of the coupling are clear,
future models may include a more realistic spatial distribution of
the physical variables.

For the study of the coupling between different oscillations we use
the perturbative method i.e. all variables are perturbed around
their equilibrium values and their consequent evolution is studied.

We first briefly describe the linear evolution of the perturbations
in form of linear waves and then study the weakly non-linear
interaction between the waves, which leads to the coupling of our
interest.

\section{The linear approximation}

All physical variables are presented as a sum of unperturbed and
perturbed parts $\xi = \xi_0 + \xi_1$. In the linear approximation
only the first order terms are retained. This  allows us to consider
the evolution of different waves separately. The system (1)-(4)
includes Alfv{\'e}n and magnetoacoustic waves.

\subsection{Torsional Alfv{\'e}n waves}

In cylindrical coordinates the pure Alfv{\'e}n waves are the
torsional waves, which have azimuthal velocity and magnetic field
components and represent the purely electromagnetic part of
oscillations. They are described by equations
\begin{equation}
{{\partial b_{\phi}}\over {\partial t}}=  B_0{{\partial
V_{\phi}}\over {\partial Z}},
\end{equation}
\begin{equation}
{\rho_0}{{\partial V_{\phi}}\over {\partial t}} =  {{B_0}\over
{\mu}}{{\partial b_{\phi}}\over {\partial Z}},
\end{equation}
where $V_{\phi}$, $b_{\phi}$ are the velocity and magnetic field
perturbations. As we already mentioned, the unperturbed density is
considered to be homogeneous.

The restoring force for the torsional waves is the tension of
magnetic field lines only. The waves propagate along the magnetic
field lines with the Alfv{\'e}n speed
\begin{equation}
c_A={{B_0}\over \sqrt{\mu\rho_0}}.
\end{equation}

Consequently the dispersion relation of the waves is
\begin{equation}
{{\omega_A}\over {k_A}}= c_A,
\end{equation}
where $\omega_A$ is the frequency and $k_A$ is the wave number.

The waves are strictly transversal and they do not cause a density
perturbation in the linear regime.

\subsection{Slow magnetoacoustic waves}

We consider the adiabatic process, then the remaining linearized
equations are
\begin{equation}
{{\partial b_R}\over {\partial t}}= B_0{{\partial V_R}\over
{\partial Z}},
\end{equation}
\begin{equation}
{{\partial b_Z}\over {\partial t}}= - {{B_0}\over R}{{\partial
(RV_R)}\over {\partial R}},
\end{equation}
\begin{equation}
{\rho_0}{{\partial V_R}\over {\partial t}} = - {{\partial p_1}\over
{\partial R}} + {{B_0}\over {\mu}}\left ( {{\partial b_R}\over
{\partial Z}} - {{\partial b_Z}\over {\partial R}}\right ),
\end{equation}
\begin{equation}
{\rho_0}{{\partial V_Z}\over {\partial t}} = - {{\partial p_1}\over
{\partial Z}},
\end{equation}
\begin{equation}
{{\partial \rho_1}\over {\partial t}}= -{\rho_0}\left ({1\over
R}{{\partial (RV_R)}\over {\partial R}} + {{\partial V_Z}\over
{\partial Z}} \right ),
\end{equation}
\begin{equation}
{p_1}= c^2_s{\rho_1},
\end{equation}
where $V_R$, $V_Z$, $b_R$, $b_Z$, $\rho_1$ and $p_1$ are the
perturbations of velocity, magnetic field, density and pressure
respectively; $c_s=\sqrt{\gamma p_0/\rho_0}$ is the adiabatic sound
speed.

After Fourier analysis with $\sim \exp{(-i\omega t + ik_z Z)}$ Eqs.
(9)-(14) lead to the equation
\begin{equation}
{{\partial^2 {\tilde V_R}}\over {\partial R^2}} + {1\over
R}{{\partial {\tilde V_R}}\over {\partial R}} + \left [m^2 - {1\over
R^2} \right ]{\tilde V_R}=0,
\end{equation}
where
\begin{equation}
m^2={{(\omega^2-c^2_sk^2_z)(\omega^2-c^2_Ak^2_z)}\over
{\omega^2(c^2_s + c^2_A)-c^2_Ac^2_sk^2_z}}.
\end{equation}
Eq. (15) is the Bessel equation when $m^2>0$, the modified Bessel
equation when $m^2<0$ and Euler equation when $m=0$ \citep{abr}.
Here $m$ plays the role of the radial wave number. For $m{\not=}0$
the expression (16) leads to the dispersion relation of obliquely
propagating magnetoacoustic waves. For the parallel propagation
(i.e. $m=0$), the waves propagate either with sound or with
Alfv{\'e}n speeds; for the high $\beta$ plasma ($c_s \gg v_A$) the
fast magnetoacoustic waves propagate with the sound speed
($\omega=\pm c_s k_z$), while the slow magnetoacoustic waves
propagate with the Alfv{\'e}n speed ($\omega=\pm c_A k_z$).

For $m^2>0$ the solution to Eq. (15) can be expressed in form of the
Bessel functions \citep{abr}
\begin{equation}
{\tilde V_R}=c_1 J_1(mR) + c_2 Y_1(mR),
\end{equation}
where $c_1$ and $c_2$ are constants.

For the parallel propagation ($m=0$) the solution to Eq. (15) can be
expressed as \citep{abr}
\begin{equation}
{\tilde V_R} = c_1R + {c_2\over R},
\end{equation}
where $c_1$ and $c_2$ are constants. This is the solution to slow
magnetoacoustic waves. For the fast magnetoacoustic waves $V_R=0$.
Imposing that the velocity is finite on the axis we obtain
\begin{equation}
{\tilde V_R} = c_1R.
\end{equation}
This solution is not valid for unbounded medium as it diverges for
large $R$. Thus in cylindrical coordinates for unbounded medium slow
magnetoacoustic waves can not propagate strictly along the magnetic
field (unlike to the Cartesian coordinates). However as far as we
are concerned in lower order harmonics of oscillations with the
wavelengths comparable to the stellar radius, then it is unjustified
to consider unbounded medium in radial direction. Therefore we
should give the boundary conditions at the stellar surface. Due to
the sharp density jump at the surface we may set up free boundary
conditions, for which the expression (19) is valid. Then the
velocity has, at the surface, the value ${\tilde V_R} = c_1R_0$,
where $R_0$ is the radius of the star.

For simplicity, we consider the slow magnetoacoustic waves
propagating along an unperturbed magnetic field in the medium
bounded along radial direction. Then the free boundary conditions at
the surface allow the radial components of velocity and magnetic
field $V_R,\,b_R$ to have the linear dependence on $R$ due to Eqs.
(9) and (19)
\begin{equation}
V_R=u_R{R\over R_0},\,\, b_R=h_R{R\over R_0}
\end{equation}
while $b_Z,\,V_Z,\,\rho_1$ remain independent on $R$. The wave
dispersion relation is the same as for the torsional waves. The
relation between the different variables in the slow magnetoacoustic
waves propagating along the magnetic field is as follows:
$$
\rho_1=-2i{{\rho_0 \omega}\over
{R_0(\omega^2-c^2_sk^2_z)}}u_R,
$$
\begin{equation}
V_Z=-2i {{c^2_Ak_z}\over {R_0(\omega^2-c^2_sk^2_z)}}u_R,
\end{equation}
$$
h_R=-{{B_0 k_z}\over {\omega}}u_R,\,\,\, b_Z=-2i{{B_0}\over {\omega
R_0}}u_R.
$$
It is easy to show from Eqs. (14) and (21), that the total pressure
perturbation at the surface vanishes for slow magnetoacoustic waves
in the case of high $\beta$ plasma ($c_s \gg c_A$) due to the
dispersion relation $\omega=\pm c_A k_z$. Indeed we have for the
total pressure perturbation:
$$p_1+{{B_0b_z}\over \mu} \approx 2i{{\rho_0 }\over
{R_0\omega}}u_R \left( {{\omega^2}\over {k^2_z}} - c^2_A\right
)=O(\beta^{-1}) ,
$$
therefore the boundary condition is satisfied in the zero order
approximation with respect to $\beta^{-1}$.

Thus the slow magnetoacoustic waves lead to the periodical
modulation of the Alfv{\'e}n speed due to the variation of density
and magnetic field. Therefore they may cause a nonlinear influence
on torsional Alfv{\'e}n waves. In the next section we study the
phenomenon in the weakly non-linear approximation.

\section{The non-linear coupling}

The non-linear influence of slow magnetoacoustic waves on the
dynamics of torsional Alfv{\'e}n waves can be considered by adding
non-linear terms to Eqs. (5)-(6), which then can be rewritten as
$$
{{\partial b_{\phi}}\over {\partial t}}=  B_0{{\partial
V_{\phi}}\over {\partial Z}} + {{\partial }\over {\partial
Z}}(V_{\phi}b_Z) - {{\partial }\over {\partial Z}}(V_Zb_{\phi})-
$$
\begin{equation}
{{\partial }\over {\partial R}}(V_Rb_{\phi} - V_{\phi}b_R),
\end{equation}
$$
({\rho_0} + \rho_1)\left ({{\partial V_{\phi}}\over {\partial t}} +
{{V_{R}}\over {R}}{{\partial (RV_{\phi})}\over {\partial R}}
+V_Z{{\partial V_{\phi}}\over {\partial Z}}  \right )=
$$
\begin{equation}
{{(B_0 + b_Z)}\over {\mu}}{{\partial b_{\phi}}\over {\partial Z}} +
{{b_R}\over {R\mu}}{{\partial (Rb_{\phi})}\over {\partial R}}.
\end{equation}

It is clear that the action has a parametric origin, because all
non-linear terms include at least one component of the torsional
waves. Here we assume that the energy of slow magnetoacoustic waves
is much larger than the energy of initial torsional waves. Then the
back-reaction of torsional waves can be neglected at the initial
stage and the slow magnetoacoustic waves can be considered as
pumping waves with constant amplitudes.

Recently \citet{zaq4} suggested that sound and linearly polarized
Alfv{\'e}n waves are coupled when they propagate with the same speed
along the magnetic field (i.e. $\beta \sim 1$) and the frequency (as
well as the wave number) of the sound waves is twice the frequency
of Alfv{\'e}n waves. In that case the action of the sound waves had
a similar parametric origin. So a similar phenomenon can occur here,
as in the high plasma $\beta$ limit the phase speeds of slow
magnetoacoustic and torsional Alfv{\'e}n waves have similar values.

In equations (5)-(6) the solution for torsional modes has an
arbitrary $R$-dependence. Therefore in the case of parallel
propagating slow waves we may take $V_{\phi}$ and $b_{\phi}$ to be
independent on $R$ in equations (22)-(23), which results in
$$
{{\partial b_{\phi}}\over {\partial t}}=  B_0{{\partial
V_{\phi}}\over {\partial Z}} + {{\partial }\over {\partial
Z}}(V_{\phi}b_Z) - {{\partial }\over {\partial Z}}(V_Zb_{\phi})-
$$
\begin{equation}
{{u_Rb_{\phi}}\over R_0} + {{h_RV_{\phi}}\over R_0},
\end{equation}
$$
({\rho_0} + \rho_1)\left ({{\partial V_{\phi}}\over {\partial t}} +
{{u_{R}V_{\phi}}\over {R_0}} +V_Z{{\partial V_{\phi}}\over {\partial
Z}}  \right )=
$$
\begin{equation}
{{(B_0 + b_Z)}\over {\mu}}{{\partial b_{\phi}}\over {\partial Z}} +
{{h_Rb_{\phi}}\over {R_0\mu}}.
\end{equation}
The physics of wave coupling will be similar for obliquely
propagating slow magnetoacoustic waves as well. In this case the
influence of slow waves on the torsional waves can be considered at
any cylindrical surface located at different $R$. Also the phase
speed of slow waves along the magnetic field coincides with the
Alfv{\'e}n speed, thus the resonant conditions given in \citet{zaq4}
will be satisfied. But, for simplicity, here we consider that both
slow and torsional waves propagate along the magnetic field.

In order to solve equations (24)-(25) we use the formalism of weakly
non-linear interaction similar to \citet{zaq4}. So we consider the
perturbations much smaller than the unperturbed values. Then the
variation of the torsional wave amplitudes due to the action of the
slow waves will be a slow process. Therefore the perturbed physical
quantities of torsional waves can be represented as the product of a
slowly varying amplitude and a rapidly oscillating term,
\begin{equation}
V_{\phi}={\hat u_{\phi}}(t)e^{i\psi_1(t,Z)} + {\hat
u^{*}_{\phi}}(t)e^{-i\psi_1(t,Z)},
\end{equation}
\begin{equation}
b_{\phi}={\hat b_{\phi}}(t)e^{i\psi_1(t,Z)} + {\hat
b^{*}_{\phi}}(t)e^{-i\psi_1(t,Z)},
\end{equation}
while the amplitude of slow waves remains constant
\begin{equation}
\rho_1={\hat \rho_1}e^{i\psi_2(t,Z)} + {\hat
\rho^{*}_1}e^{-i\psi_2(t,Z)},
\end{equation}
\begin{equation}
V_Z={\hat u_Z}e^{i\psi_2(t,Z)} + {\hat u^{*}_Z}e^{-i\psi_2(t,Z)},
\end{equation}
\begin{equation}
u_R={\hat u_R}e^{i\psi_2(t,Z)} + {\hat u^{*}_R}e^{-i\psi_2(t,Z)},
\end{equation}
\begin{equation}
b_R={\hat b_R}e^{i\psi_2(t,Z)} + {\hat b^{*}_R}e^{-i\psi_2(t,Z)}.
\end{equation}
Here
\begin{equation}
{\psi_1}=-{\omega_A}t + k_A Z,
\end{equation}
\begin{equation}
{\psi_2}=-{\omega}t + k_z Z,
\end{equation}
where ${\omega}$, $k_z$, ${\omega_A}$, $k_A$ are the frequencies and
wave numbers of slow magnetoacoustic and torsional Alfv{\'e}n waves
respectively. The frequencies and wave numbers satisfy the
dispersion relation (8) which is the same for both waves.

Substitution of expressions (26)-(31) into equations (24)-(25), and
averaging over rapid oscillations in phase, leads to the
disappearance of all exponential terms, so that only the first and
third order terms remain. In the first approximation (neglecting all
second and third order terms), the slow magnetoacoustic and
torsional Alfv{\'e}n waves are decoupled and the amplitudes are
constant. Third order terms (which are due to the advective terms in
the momentum equation) are significant only in the case of very
large amplitudes and so can be ignored. However, when
\begin{equation}
\psi_2 = {\pm}2\psi_1,
\end{equation}
the second order terms also remain after averaging and consequently
the torsional wave amplitudes become time-dependent. So if the
frequencies and wave numbers satisfy the conditions
\begin{equation}
\omega=2\omega_A,~ k_z=2k_A,
\end{equation}
\begin{equation}
\omega=-2\omega_A,~ k_z=-2k_A,
\end{equation}
the waves are resonantly coupled and the slow magnetoacoustic waves
may transfer energy to torsional Alfv{\'e}n waves.

According to condition (35), averaging equations (24)-(25) leads to
the equations which govern the temporal behaviour of the complex
amplitudes:

\begin{equation}
{{\partial {\hat b_{\phi}}}\over {\partial t}} -i\omega_A {\hat
b_{\phi}}+ i\omega_A {{{\hat \rho_1}}\over {\rho_0}}{\hat
b^*_{\phi}}-ik_A B_0{\hat u_{\phi}}=0,
\end{equation}
$$
{{\partial {\hat u_{\phi}}}\over {\partial t}}+{{{\hat \rho_1}}\over
{\rho_0}}{{\partial {\hat u^*_{\phi}}}\over {\partial t}} -i\omega_A
{\hat u_{\phi}}+{{2{\hat u_R}}\over R_0}{\hat u^*_{\phi}}-{{ik_A
B_0}\over {\mu \rho_0}}{\hat b_{\phi}}+
$$
\begin{equation}
{{2k_A B_0}\over {\mu \rho_0 \omega_A R_0}}{\hat u_R}{\hat
b^*_{\phi}}=0.
\end{equation}
Here ${\hat \rho_1}={\hat \rho_{11}}+i {\hat \rho_{12}}$ and ${\hat
u_R}={\hat u_{R1}}+i{\hat u_{R2}}$ are constants.

Now let us search for solutions of (37)-(38) in the exponential form

\begin{equation}
{\hat b_{\phi}}=({\hat b_{\phi 1}}+i{\hat b_{\phi 2}})e^{{\delta}t},
\end{equation}
\begin{equation}
{\hat u_{\phi}}=({\hat u_{\phi 1}}+i{\hat u_{\phi 2}})e^{{\delta}t},
\end{equation}
where ${\hat b_{\phi 1}}$, ${\hat b_{\phi 2}}$, ${\hat u_{\phi 1}}$
and ${\hat u_{\phi 2}}$ are constants. In expressions (39)-(40)
$\delta$ is real, otherwise equations (37)-(38) are not satisfied. A
positive $\delta$ implies amplification of the initial perturbation,
while a negative $\delta$ implies damping.

We substitute expressions (39)-(40) into equations (37)-(38) and,
without loss of generality, assume ${\hat \rho_{11}}={\hat
u_{R2}}=0$. Finally, we get equations
\begin{equation}
\left ( \delta -\omega_A{{{\hat \rho_{12}}}\over {\rho_0}} \right
){\hat b_{\phi 1}}+\omega_A {\hat b_{\phi 2}} +k_AB_0{\hat u_{\phi
2}}=0,
\end{equation}
\begin{equation}
\left ( \delta +\omega_A{{{\hat \rho_{12}}}\over {\rho_0}} \right
){\hat b_{\phi 2}}-\omega_A {\hat b_{\phi 1}} -k_AB_0{\hat u_{\phi
1}}=0,
\end{equation}
$$
\left ( \delta +{{2{\hat u_{R1}}}\over {R_0}} \right ){\hat u_{\phi
1}}+\left ( \delta {{{\hat \rho_{12}}}\over {\rho_0}} +\omega_A
\right ){\hat u_{\phi 2}}+
$$
\begin{equation}
{{2k_A B_0}\over {\mu \rho_0 \omega_A R_0}}{\hat u_{R1}}{\hat
b_{\phi 1}} + {{k_A B_0}\over {\mu \rho_0}}{\hat b_{\phi 2}}=0,
\end{equation}
$$
\left ( \delta -{{2{\hat u_{R1}}}\over {R_0}} \right ){\hat u_{\phi
2}}+\left ( \delta {{{\hat \rho_{12}}}\over {\rho_0}} -\omega_A
\right ){\hat u_{\phi 1}}-
$$
\begin{equation}
{{k_A B_0}\over {\mu \rho_0}}{\hat b_{\phi 1}}- {{2k_A B_0}\over
{\mu \rho_0 \omega_A R_0}}{\hat u_{R1}}{\hat b_{\phi 2}}=0.
\end{equation}

This system has a non-trivial solution if the system determinant is
zero, which gives a fourth order equation for $\delta$. After
dropping higher order terms, we get
\begin{equation}
{{\delta}}\approx {\omega_A \over 2}{{{\hat \rho_{12}}}\over
{\rho_0}}.
\end{equation}
Thus the torsional Alfv{\'e}n waves grow exponentially in time and
their growth rate depends on the relative amplitude of slow
magnetoacoustic waves i.e. ${{{\hat \rho_{12}}}/{\rho_0}}$, which is
typical of weakly non-linear processes.

In order to check the analytical solution, we performed the
numerical simulation of spatially averaged equations (24)-(25) with
$k_z=2k_A$, so that only the time dependence is retained. Then the
numerical calculation is straightforward and we find that torsional
waves have an exponentially growing solution when their period is
twice the period of the slow magnetoacoustic waves i.e.
$\omega=2\omega_A$ (Fig.2). The up-plots refer to the density and
velocity components of the slow waves ($\rho_1$ and $u_R$) and the
down-plots refer to the components of the torsional waves ($b_\phi$
and $u_\phi$). Here $\beta$ is assumed to be $\sim$ 10$^2$ and the
strong amplification of torsional waves begins after $\sim$ 10-15
wave periods.

Note that the growth rate of torsional Alfv{\'e}n waves depends on
the value of plasma $\beta$. The energy transfer occurs only when
the density perturbations in slow magnetoacoustic waves are nonzero.
When $\beta$ goes to infinity, then the density perturbations vanish
and so there is no energy exchange between the waves. Thus the
resonant process is of importance for large, but finite $\beta$,
which requires strong magnetic fields in the stellar interiors.

\section{Discussion}

We have seen that slow magnetoacoustic and torsional Alfv{\'e}n
waves are resonantly coupled when they propagate with the same
Alfv{\'e}n speed along an unperturbed magnetic field in a $\beta
\gg1$ medium. The slow magnetoacoustic wave parametrically triggers
the torsional wave with the half frequency and wave number. The
growth rate of the torsional mode depends on the amplitude of the
slow magnetoacoustic wave: the larger is the amplitude of the slow
magnetoacoustic wave, the faster is the energy transfer to torsional
waves, which is a common scenario for non-linear processes. The
growth rate depends on the value of plasma $\beta$ as well; higher
$\beta$ leads to weaker coupling. The reason is that the density
perturbations in slow magnetoacoustic waves are weaker for higher
$\beta$ and thus the slow waves do not alter the Alfv{\'e}n speed.
However the situation probably will be changed for nonadiabatic slow
waves as the density perturbations can be larger due to the
compensation by the temperature variation. Then the density
perturbations will affect significantly the Alfv{\'e}n speed, thus
the torsional waves. This process seems to have interesting
consequences and will be studied in future.

The coupling between these waves may have interesting applications
in astrophysics. The existence of torsional oscillations in stellar
interiors is a long time dilemma. Hydromagnetic torsional
oscillations of a seed magnetic field have been suggested to take
place in stellar interiors, which may lead to the cyclic change in
sign of the solar toroidal magnetic field \citep{wal,cow,pid,go}.
However the oscillations probably are damped because of the
transport of the magnetic flux to the surface and/or the phase
mixing \citep{cha}. Therefore the absence of a plausible energy
source which can support torsional oscillations has been an argument
against their existence \citep{ros}. However, several different
sources have previously been suggested to support torsional
oscillations \citep{la,ma,lan,ga}, but they are not convincing.

Recently stellar radial pulsation in the fundamental mode, which can
be considered as a standing fast magnetoacoustic wave in the
presence of magnetic field, has been suggested as a source for
torsional Alfv{\'e}n waves \citep{zaq3}. However, amplified
torsional waves can not give rise to torsional oscillations due to
their short wavelength as compared to the stellar radius (because
the Alfv{\'e}n speed is much smaller than the sound speed).
Therefore, rather than give rise to large-scale torsional
oscillations, they will propagate upwards to the stellar atmosphere,
and can be observed in the stellar wind \citep{zaq5}.

It is a natural physical process that in closed systems (the Sun or
a simple tuning fork) almost all energy is stored in the fundamental
mode. For example, take an ordinary tuning fork: any force (whatever
its origin is) leads to an acoustic oscillation in the fundamental
frequency. Obviously, there are other overtones, but with much
smaller amplitudes. However, if one puts the tuning fork in a
magnetic field embedding a plasma, then the acoustic frequency will
split into two frequencies corresponding to fast and slow
magnetoacoustic waves. Similarly, the large-scale magnetic field may
cause the splitting of the fundamental acoustic mode into high and
low frequency branches, corresponding to fast and slow
magnetoacoustic waves. Therefore a small deviation of the star from
equilibrium must lead to the oscillation in both, fast and slow
fundamental modes. Due to the small phase velocity, the period of
standing slow magnetoacoustic waves will be much longer than the
period of the fast fundamental mode. In the case of the adopted
simple configuration of poloidal magnetic field in a stellar
interior (see Fig.1), the standing slow magnetoacoustic waves can be
pictured as radial oscillations with velocity nodes along the $Z$
axis, which causes a periodic variation in stellar oblateness The
number of nodes depends on which harmonic is excited. As the phase
speed is Alfv{\'e}nic, the first harmonic will have a period of
order $2R_0/c_A$, the second of order $R_0/c_A$ etc. Due to the
radial polarization of the velocity, the oscillations may easily
gain energy from various sources like gravity, pressure gradient,
rotation etc.

Because of the parametric coupling, these oscillations may transfer
energy to the torsional Alfv{\'e}n waves with twice the period and
wavelength (see equation (35) and Fig.2). The amplified torsional
Alfv{\'e}n waves  may easily lead to the set up of torsional
oscillations as their wavelength now is comparable to the stellar
radius. Torsional oscillations now can be sustained against damping,
as they may continuously gain energy from the radial pulsation.

The energy that may support the slow magnetoacoustic fundamental
mode can be large. With reference to subsection 5.1., the
gravitational energy of the Sun (i.e. the work required to dissipate
solar matter to infinity), with radius $R_0\approx 7{\cdot}10^{10}$
cm and mass $M\approx 2{\cdot}10^{33}$ g, is $6.6{\cdot}10^{48}$
erg, even larger than the total internal radiant energy $\approx
2.8{\cdot}10^{47}$ erg \citep{al}. The global density variation due
to the slow magnetoacoustic fundamental mode is accompanied by a
gravitational potential variation (see the classical paper by
\citet{led}), which unfortunately is neglected in this paper due to
the simplicity of presentation. Suppose that the radius variation
$\Delta R$ due to the slow magnetoacoustic fundamental mode is as
small as $\Delta R/R_0 \sim 10^{-5}$ (for the Sun it is $\sim
0.01^{\prime\prime}$, however observations also show radius
variations $\sim 0.02^{\prime\prime}$ \citep{em} and even $\sim
0.08^{\prime\prime}$ \citep{lac}). The gravitational energy change
for a homogeneous sphere \citep{chan} will be of the order of
$6.6{\cdot}10^{48} \Delta R/R_0 \sim 6.6{\cdot}10^{43}$ erg (this is
the energy stored in the fundamental mode). Suppose that only
$0.1\%$ of this energy goes into the energy of torsional
oscillations. Then the corresponding magnetic energy variation will
be of the order of $\sim 6.6{\cdot}10^{40}$ erg. Even if the
magnetic field is redistributed throughout the whole stellar volume,
we get a toroidal magnetic field variation with an amplitude
$>10^{4}$ G. So the energy is more than enough for supporting the
torsional oscillation against ohmic diffusion or phase mixing
\citep{cha}. Moreover, the amplification of the toroidal magnetic
field component will be accompanied by the magnetic buoyancy
instability, which leads to the eruption of magnetic flux towards
the surface. Thus torsional oscillations can not cause a significant
back-reaction on the pumping radial pulsation as their further
amplification will be balanced by the energy transport towards the
surface in form of magnetic flux tubes, which leads to chromospheric
and coronal activity.

However, it must be mentioned that the many simplifications of the
model (e.g. uniform density, cylindrical geometry) will cause
significant limitations to the application of the mechanism to real
situations in stars. For instance, notice that the density (and the
magnetic field) can be considered as uniform only in a narrow
cylindrical shell, not throughout the star, but the physics of the
mechanism still remains substantially the same. Considering
spherical symmetry instead of cylindrical one will be a further
improvement, however it will cause additional mathematical
complications. The aim of this paper is to show the existence of
such an energy transformation mechanism, which can be developed and
applied to more realistic astrophysical situations in future.

Here we briefly discuss possible astrophysical applications of the
wave coupling discussed in this paper.

\subsection{Solar radius variation and torsional oscillations}

The variation of the solar radius during the 11-year period activity
cycle is a long time problem. The characteristics of the variation
may reveal some basic mechanisms of activity, therefore many efforts
have been made in order to find observational evidence of a
correlation between radius variation and activity parameters. The
observational results are controversial and somewhat inconsistent.
Some observations indicate a negative correlation \citep{lac}, some
results show no systematic variation in time \citep{bro}. Most
observations reveal a positive correlation \citep{ul,ba,em,no}. The
observed amplitude of the variation is also controversial. Based on
Mount Wilson data, \citet{ul} found $\sim 0.2^{\prime\prime}$;
\citet{no} suggested a similar amplitude $\sim 0.2^{\prime\prime}$,
while in \citet{lac} the amplitude is  $\sim 0.08^{\prime\prime}$.
Probably, the most realistic value of the amplitude  $\sim
0.02^{\prime\prime}$ was obtained by \citet{em} by analyzing
SOHO-MDI data. The variation of the solar radius during the activity
cycle naturally leads to the idea of its connection to the magnetic
field pressure variation in the interior due to the large-scale
turbulent dynamo. However, even a very small variation of the radius
requires a huge magnetic energy to compensate the corresponding
gravitational energy change (see the above estimate). Then it seems
plausible that the radius variation is due to the global slow
magnetoacoustic mode. Slow magnetoacoustic waves propagating along
the poloidal magnetic field have both $R$ and $Z$-components of
velocity (see Eqs.(9)-(14)), therefore the global slow mode will
lead to the periodical variation of the solar oblateness, which may
cause interesting consequences for solar total irradiance. The
fundamental period of pulsations in the global slow magnetoacoustic
mode will be of order $2R_0/c_A$, which for the 11-year period
implies the mean magnetic field strength in the solar interior to be
$\sim 10^3$ G.

Another possible candidate for the global mode coupling can be the
observed 1.3 yr variation at the base of the solar convection zone
\citep{howe}. If the global slow mode has the period of $\sim$ 0.65
yr, then it may drive the torsional oscillations with the period of
1.3 yr in the solar tachocline.

\subsection{Orbital period modulation in close binaries}

Long-term observations of close binary stars show a variation of the
orbital period over long time-scales with relative amplitude
${\Delta P}/P \sim 10^{-5}$, where $P$ is the orbital period. A
variety of theoretical models have been proposed to explain this
phenomenon. The first models included the observed period changes
due to the apsidal motion, or the influence of a distant, unseen
companion. Then \citet{ap} suggested that the variation can be
explained by the gravitational coupling of the orbit to the
variation in shape of a magnetically active companion. The variable
deformation of the active star is produced by the variation in the
distribution of angular momentum during the activity cycle. On
another hand, \citet{lan} suggested that a torsional oscillation
driven by the stellar dynamo may account for the observed behaviour
of the binary system. However the problem still remains, because the
magnetic energy can not compensate the gravitational energy change,
as has been shown for the Sun.

Instead, here we propose that the long term radial pulsation due to
the global slow mode  considered in our paper (which causes a
periodic variation in oblateness) may be responsible for the
variation of the orbital period with a gravitational coupling
similar to Applegate's. However contrary to Applegate's mechanism,
here stellar radial pulsation can be the reason for both orbital
period modulation and magnetic activity in the primary star.

The relative amplitude of the orbital period modulation is
proportional to the relative variation of the gravitational
acceleration of the primary ${\Delta P}/P\sim {\Delta g}/g$
\citep{ap}. On the other hand, the relative variation of the
gravitational acceleration is of the order of ${\Delta g}/g\sim
{\Delta R}/R$, where $R$ is radius of the primary. Then in order to
get the observed relative period modulation ${\Delta P}/P\sim
10^{-5}$, the radius of primary needs to vary as much as ${\Delta
R}/R\sim 10^{-5}$. It is very intriguing that the observed solar
radius relative variation deduced from SOHO data \citep{em} has a
comparable extent ${\Delta R}/R_0\sim 2{\cdot}10^{-5}$. So relative
radius variations of the primary with amplitude similar to the
Sun's, can explain the observed orbital period modulation in binary
systems.

\section{Conclusion}

In this paper we considered the weakly non-linear coupling between
slow magnetoacoustic and torsional Alfv{\'e}n waves in the high
plasma $\beta$ limit. It has been shown that slow magnetoacoustic
waves propagating along the unperturbed magnetic field may
resonantly amplify torsional Alfv{\'e}n waves with the half
frequency and wave-number of slow magnetoacoustic waves. The growth
rate of torsional Alfv{\'e}n waves depends on the amplitude of slow
magnetoacoustic waves and the value of plasma $\beta$. Therefore the
global slow magnetoacoustic waves, excited due to the action of
various forces (gravity, gas and radiation pressure gradients,
rotation etc.) in a stellar interior, may give rise to large-scale
torsional oscillations. The process may be of importance in solar
and stellar physics, but needs more detailed study.



\acknowledgments We thank an anonymous referee for constructive
criticism and useful suggestions. The work was supported by the
grant of Georgian National Science Foundation GNSF/ST06/4-098 and
the NATO Reintegration Grant FEL.RIG 980755. This work was carried
out during a visit of T.Z. to the Dipartimento di Fisica e
Astronomia, Sezione Astrofisica, Universit{\'a} di Catania.

\clearpage



\begin{figure}
   \centering
   \includegraphics[width=6cm]{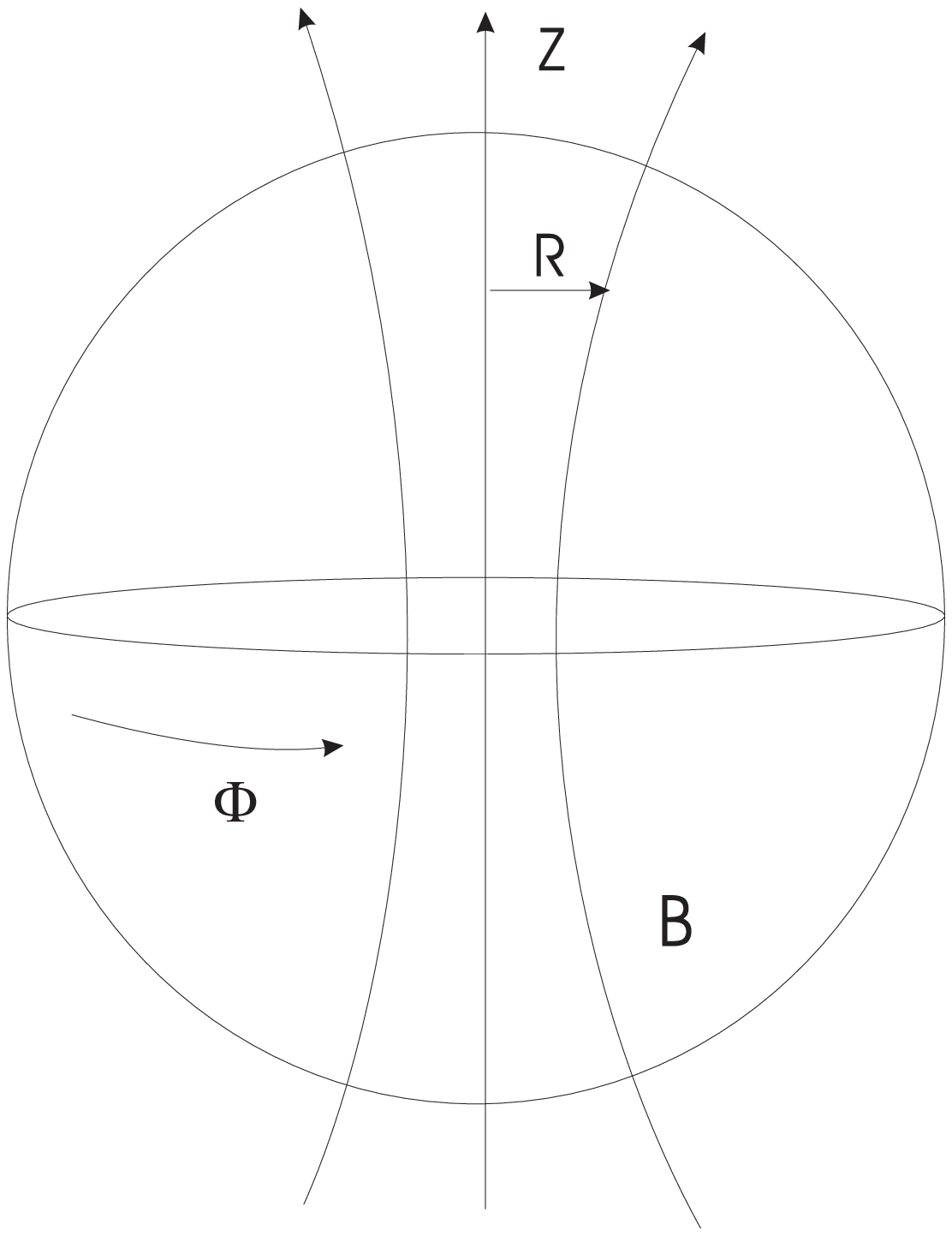}
      \caption{The cylindrical coordinate system and poloidal magnetic
field in stellar interior.
              }
         \label{FigVibStab}
   \end{figure}

\clearpage


\begin{figure}
   \centering
   \includegraphics[width=14cm]{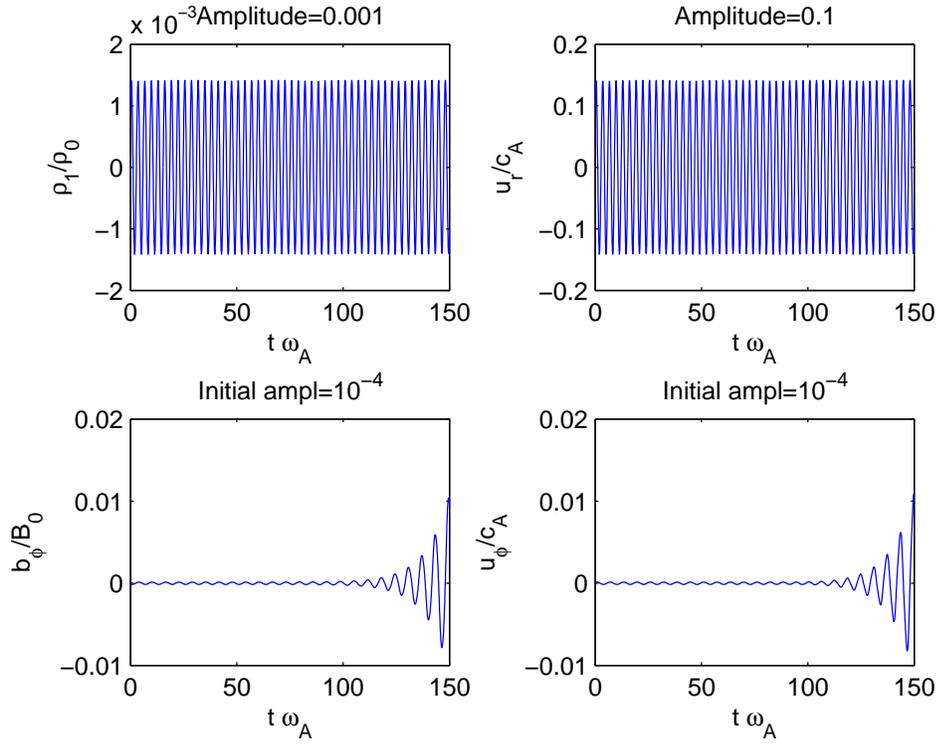}
      \caption{The time evolution of spatially averaged (with $Z$) slow magnetoacoustic (up) and
torsional modes (down). The strong amplification of the torsional
mode with twice the period of the slow magnetoacoustic mode is
clearly seen, as predicted by the analytical solution. Here
$\rho_1/\rho_0=0.001$ is considered.
              }
         \label{FigVibStab}
   \end{figure}








\end{document}